\documentclass[english,prd,twocolumn,superscriptaddress,nofootinbib,preprintnumbers]{revtex4}
\usepackage[latin1]{inputenc}
\usepackage{graphicx}
\usepackage{amssymb}
\usepackage{color}
\usepackage{float}
\usepackage{amsmath}
\usepackage{amsfonts}
\usepackage{dcolumn}
\usepackage{hyperref}

\newcommand{\be}{\begin{eqnarray}}
\newcommand{\ee}{\end{eqnarray}}
\newcommand{\ba}{\begin{eqnarray}}
\newcommand{\ea}{\end{eqnarray}}

\def\ben{\begin{equation}}
\def\een{\end{equation}}
\def\bena{\begin{eqnarray}}
\def\eena{\end{eqnarray}}

\begin{document}

\title{Late time acceleration in a slow moving galileon field}

\author{Debabrata Adak}
\author{Amna Ali}
\author{Debasish Majumdar}
\affiliation{Astroparticle Physics and Cosmology Division and Centre
for Astroparticle Physics, Saha Institute of Nuclear Physics, 1/AF
Bidhannagar, Kolkata 700 064, India}

\begin{abstract}

In this paper, we examine the cosmological viability of
 a slow moving galileon field in a potential. The Lagrangian $\cal{L}$ $ =-\frac{1}{2}g^{\mu\nu}\pi_{;\mu}\pi_{;\nu}
+\frac{G^{\mu\nu}}{2M^2}\pi_{;\mu}\pi_{;\nu}$ respects the galileon symmetry in curved space time. 
 We carry out detailed investigations of the underlying dynamics of this Lagrangian with Einstein-Hilbert term
and a potential. We demonstrate that the model can give rise to
 a viable ghost free late time acceleration of universe. Furthermore we study the cosmological perturbation of the model
and see that the model gives different BBN constraints at early times. We also carry out the observational analysis of the model and 
use observational data from growth, Type
Ia Supernovae (SNIa), Baryon Acoustic Oscillations (BAO) and Cosmic Microwave Background
(CMB) to constrain the parameters of the theory. 
\end{abstract}

\maketitle

\section{Introduction}
From cosmological observations \cite{Perlmutter:1998np,Riess:1998cb,Seljak:2004xh,Spergel:2006hy} it is evident
 that our universe is currently undergoing an accelerated expansion. The theoretical understanding of the nature
 of cosmic repulsion  is a challenge for the cosmologist and the particle physicist today. A variety of 
approaches have been studied to address the problem, still till date there is no definite clue for it.
 According to the standard approach, the late time acceleration can be accounted for by supplementing 
the energy momentum tensor by an exotic fluid component with large negative pressure dubbed
 {\it dark energy} \cite{review1,vpaddy,review2,review3,review3C,review3d,review4,review5}, which constitutes 
of about three fourth of total cosmic budget of the universe \cite{planck}. The simplest candidate of dark energy
 which is consistent with the observations is provided by cosmological constant $\Lambda$. However, there are various
 serious theoretical problems associated to it namely, the {\it  fine tuning} and the {\it coincidence} problem.
 
An interesting alternative to cosmological
constant is provided by the scalar fields. The cosmological dynamics of a variety of
scalar field models has been studied in the literature \cite{review2}. Though these models do not address
 the cosmological constant problem, they can give late time cosmic acceleration  and can also provide a viable cosmological
dynamics at early epochs. Scalar field models with generic features like the trackers
are capable of alleviating the fine tuning and coincidence problems.
At present, these scalar field models are absolutely consistent with
the observations but at the same time, a large number of these
models are also permitted. One must therefore rely on  future data  which should allow to narrow down
the class of permissible scalar field dark energy models. 

Another interesting approach employed to explain the late time acceleration of the 
universe is to modify the gravity at large scale (infra-red modification of gravity). It is well known that gravity gets  quantum mechanically corrected at small scales which at present is beyond our
 observational reach, therefore it might be possible that gravity also suffers modifications at large scales, where it is never tested directly. The modified gravity models
are either phenomenological \cite{fr} or are motivated by the higher dimensions \cite{galileon}. However, any large
scale modification of gravity should be capable of being
distinguished from cosmological constant, should be free from ghost and 
tachyon instabilities and  should not conflict with the local physics.
A large number of modified gravity models have been investigated,
 among which is the galileon gravity \cite{galileon}.  It is motivated by the decoupling 
limit of the Dvali-Gabadadze-Porrati (DGP) model \cite{dgp}. Galileon theories are subclass of the scalar-tensor theories involving only up to second order derivatives, which was
 originally found by Hordenski \cite{hordenski}. The Lagrangian of the galileon field $\pi$  respects the shift 
or the galileon symmetry in flat space time $\pi\rightarrow \pi + a + b_\mu x^\mu $, where
 $a$ and $b_\mu$ are a constant and a constant vector respectively. Due to this symmetry, the equations of motion for 
the field contain only second derivatives. The galileon modified gravity can give rise to late time acceleration and
is free from negative energy instabilities \cite{gannouji1}. Galileon field in a potential is also 
studied and is shown to give a viable cosmological dynamics \cite{amna:2012}.

 Recently the galileon symmetry was subsequently extended to the curved space time by the authors of \cite{Slow1}, and was 
shown that the Lagrangian $\cal{L}$ $ =-\frac{1}{2}g^{\mu\nu}\pi_{;\mu}\pi_{;\nu}+\frac{G^{\mu\nu}}{2M^2}\pi_{;\mu}\pi_{;\nu}$ 
respects this symmetry in curved space time. The sign of the terms in $\cal{L}$ are chosen in such 
a way that, the effective propagator of $\pi$ is never ghost-like and hence are stable.
By adding the standard Einstein-Hilbert term to $\cal{L}$, 
and a non trivial potential for $\pi$, one gets a simple
though rich gravitational theory, with some nice properties. In particular, in the flat space time limit and  in the
 regimes in which the analogue of the strong energy condition is violated, 
the field $\pi$ moves slower than in the cousin canonical theory. For this reason, $\pi$
is dubbed as the ``Slotheon''.

In this paper we investigate the cosmological dynamics of a model based upon slotheon gravity, set up the autonomous
system and discuss the existence and stability of fixed
points. We study the accelerating solution and the observational constraints on the
model parameters using, growth, supernovae, BAO and CMB
data. We also study the metric perturbations and investigate the growth history of the model.

\section{ Background}
In slotheon theories, the large scale modification of
gravity arises due to self interaction of a scalar field $\pi$, which moves in a potential $V$ and
couples with matter and metric. In what follows, we shall
consider the slotheon action of the form,

\begin{align}
S=\int d^4x \sqrt{-g}\Bigl[\frac{1}{2}\Bigl(M_{\rm {pl}}^2 R &-\Bigl(g^{\mu\nu} -\frac{G^{\mu\nu}}{M^2}\Bigr)\pi_{;\mu}\pi_{;\nu}\Bigr)-V(\pi)\Bigr]\nonumber\\
&+ \mathcal{S}_m\Bigl[\psi_m;e^{2 \beta \pi/M_{\rm pl}} g_{\mu\nu}\Bigr]\,,
\label{1.1}
\end{align}
where $M_{\rm{pl}}^2=\frac{1}{8\pi G}$ is the reduced Planck mass, $M$ is a
energy scale, R is the Ricci Scalar, $\psi_m$ is the matter field which couples to $\pi $ 
and $\beta$ is dimensionless coupling constant. Variation of this action gives the following equations of motion

\begin{align}
M^2_{\rm{pl}} G_{\mu\nu}= T_{\mu\nu}^{(m)}+T_{\mu\nu}^{(r)}+T_{\mu\nu}^{(\pi)}\,,
\label{1.2}
\end{align}
\begin{align}
\Box \pi+\frac{1}{M^2}\Bigl[\frac{R}{2}\Box \pi-R^{\mu\nu}\pi_{;\mu\nu}\Bigr]-V'(\pi)=-\frac{\beta}{M_{\rm{pl}}}T^{(m)}\,,
\label{1.3}
\end{align}
where $T_{\mu\nu}^{(m)}, T_{\mu\nu}^{(r)}, T_{\mu\nu}^{(\pi)}$ corresponds to the energy momentum tensor of dust like 
particles, radiation and field respectively, and
\begin{align}
T_{\mu\nu}^{(\pi)}&=
\pi_{;\mu}\pi_{;\nu}-\frac{1}{2}g_{\mu\nu}(\nabla\pi)^2
-g_{\mu\nu}V(\pi)\nonumber\\
&+\frac{1}{M^2} \Bigl[\frac{1}{2}\pi_{;\mu}\pi_{;\nu}R-2\pi_{;\alpha}\pi_{(;\mu}R^{\alpha}_{\nu)}+\frac{1}{2}\pi_{;\alpha}\pi^{;\alpha}G_{\mu\nu}\nonumber\\
&-\pi^{;\alpha}\pi^{;\beta}R_{\mu\alpha\nu\beta}-\pi_{;\alpha\mu}\pi^{\alpha}_{;\nu}+\pi_{;\mu\nu}\pi_{;\alpha}^{~\alpha}\nonumber\\
&+\frac{1}{2}g_{\mu\nu}[\pi_{;\alpha\beta}\pi^{;\alpha\beta}-(\pi_{;\alpha}^{~\alpha})^2+2\pi_{;\alpha}\pi_{;\beta} R^{\alpha\beta}]\Bigr]\,,
\end{align}
where $'$ denotes the derivative wrt $\pi$. It has been shown that, for $V(\pi)>0$, the time derivative of the $\pi$ field is smaller than that of a canonical scalar field with the same energy density. 
Therefore  the $\pi$ field  is slower than a canonical scalar field \cite{Slow1, Slow2}. Though due to the presence of 
potential the action is not $\pi$-parity invariant, yet it is free from 
Ostrogradsky ghost problem. Also the slowing of the field $\pi$ is solely due to gravitational interaction.
 In a spatially flat  FLRW background, the equations of motion take the form

\begin{align}
3M_{\rm{pl}}^2H^2 &=\rho_m+\rho_r+\frac{\dot{\pi}^2}{2}+\frac{9H^2\dot{\pi}^2}{2M^2}+V{(\pi)}\,,\\
M_{\rm{pl}}^2(2\dot H + 3H^2)&=-\frac{\rho_r}{3}-\frac{\dot{\pi}^2}{2}+V(\pi)+\frac{\dot{\pi}^2}{2M^2}\Bigl(2\dot H + 3H^2\Bigr)\nonumber\\
&+\frac{2H\dot{\pi}\ddot{\pi}}{M^2}\,,\\
-\frac{\beta}{M_{\rm{pl}}} \rho_m &=\ddot{\pi}+3H\dot{\pi}+\frac{3H^2}{M^2}\Bigl(\ddot{\pi}+3H\dot{\pi}+\frac{2\dot{H}\dot{\pi}}{H}\Bigr)\nonumber\\
&+V'(\pi).
\end{align}

The equation for the conservation of energy, derived from the previous equations are
\begin{align}
\dot\rho_m+3H\rho_m &=\frac{\beta}{M_{\rm{pl}}}\dot{\pi} \rho_m ,\\
\dot\rho_r+4H\rho_r &=0.
\end{align}

Let us introduce the following dimensionless variables
\begin{align}
x&=\frac{\dot{\pi}}{\sqrt{6}H M_{\rm{pl}}}\,,\quad y=\frac{\sqrt{V(\pi)}}{\sqrt{3} H M_{\rm{pl}}}\,,
\label{xy} \\
\epsilon &=\frac{H^2}{2M^2}\,, \quad \lambda=-M_{\rm{pl}}\frac{V'(\pi)}{V(\pi)}\,,
\label{epsilon}
\end{align}
\begin{figure}[t!]
\centering
\includegraphics[scale=.6]{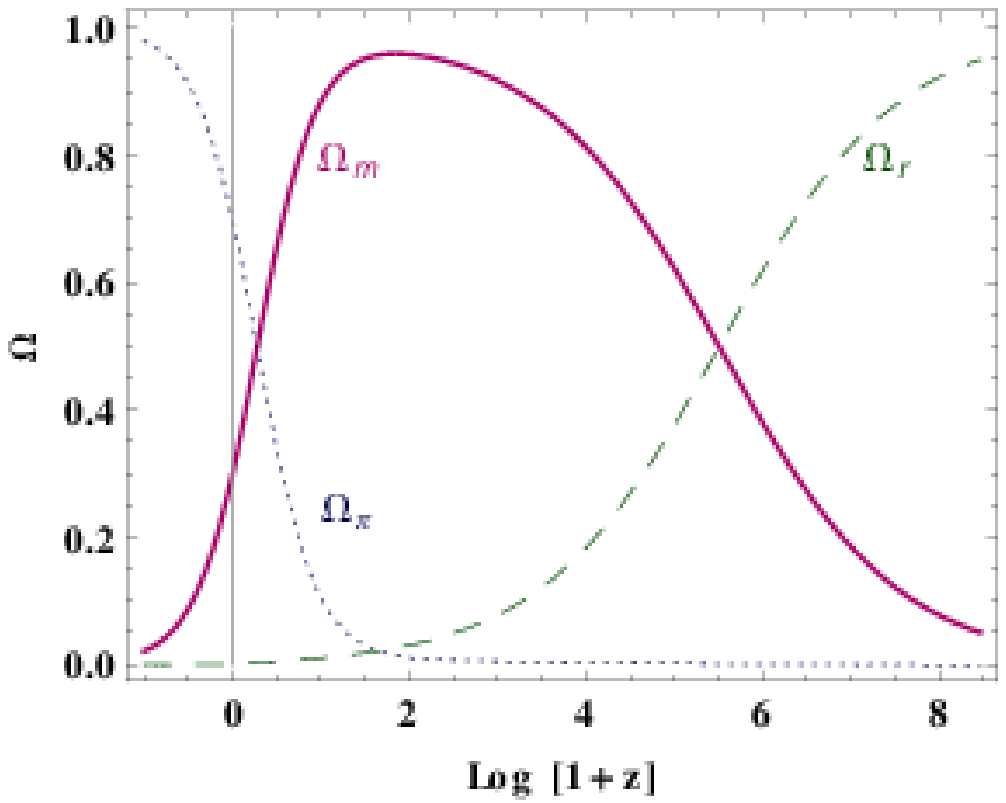}
\includegraphics[scale=.6]{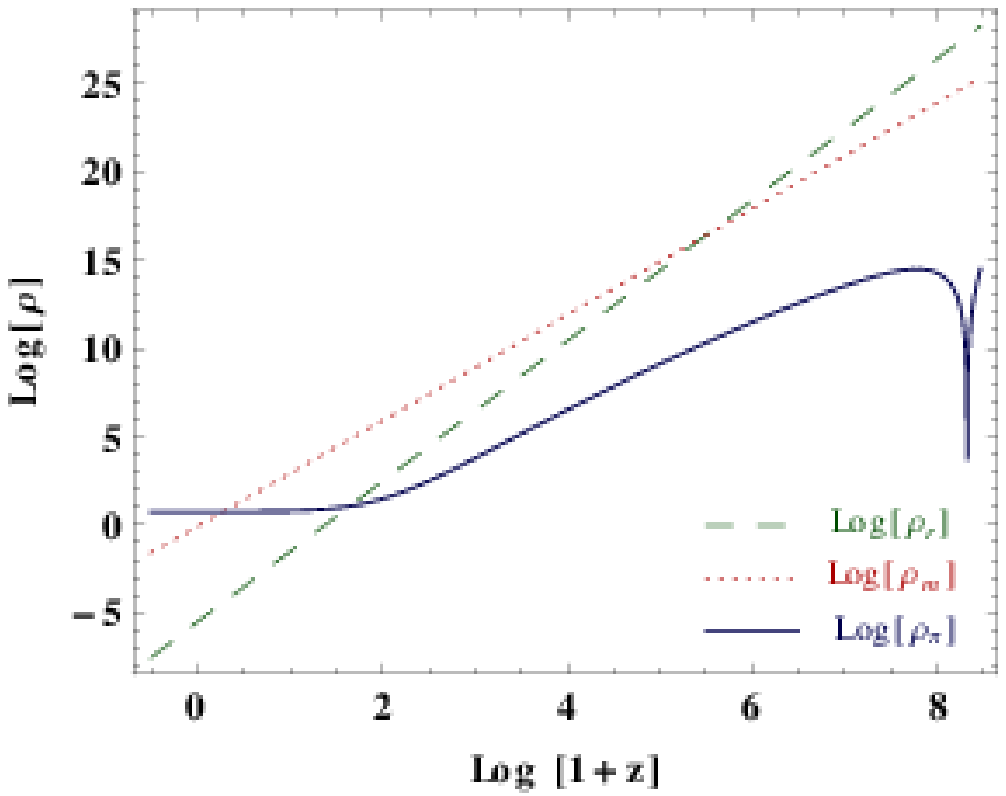}
\caption{(Top): Density parameters of matter($\Omega_m $), radiation($\Omega_r$) and field($\Omega_{\pi}$) 
for potential (\ref{pot}) are shown here with $\beta=0.1$.
 (Bottom): Density $\log\Bigl(\frac{\rho}{3M_{\rm pl}^2H_0^2}\Bigr)$ for the 3 fields are shown for the same value of $\beta$.}
\label{Fig1}
\end{figure}
\begin{figure}[h]
\centering
\includegraphics[scale=0.6]{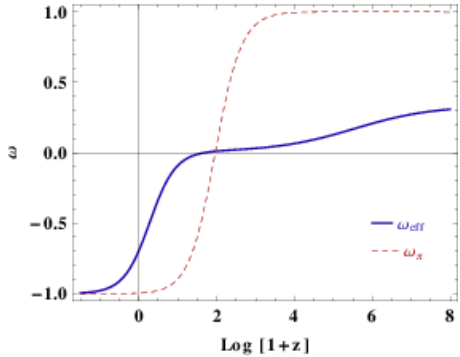}
\caption{The cosmic evolution of the field equation of state $w_\pi$ and the total effective equation of state $w_\text{eff}$ for  $\beta=0.1$.}
\label{Fig2}
\end{figure}
required to cast the evolution equations in the form of an autonomous
system
\begin{align}
\frac{{\rm d}x}{{\rm d}N}&=x\Bigl(\frac{\ddot{\pi}}{H\dot{\pi}}-\frac{\dot H}{H^2}\Bigr),\\
\frac{{\rm d}y}{{\rm d}N}&=-y \Bigl(\sqrt{\frac{3}{2}}\lambda x+\frac{\dot H}{H^2}\Bigr),\\
\frac{{\rm d}\epsilon}{{\rm d}N}&=2\epsilon\frac{\dot H}{H^2},\\
\frac{{\rm d}\Omega_r}{{\rm d}N}&=-2\Omega_r\Bigl(2+\frac{\dot H}{H^2}\Bigr),\\
\frac{{\rm d}\lambda}{{\rm d}N}&=\sqrt{6}x\lambda^2(1-\Gamma)\,,
\end{align}
where $N\equiv \ln a$, $\Gamma=\frac{VV_{,\pi\pi}}{V_{,\pi}^2}$ and

\begin{align}
\frac{\dot H}{H^2}&=\frac{-3x^2(1+24\epsilon+108\epsilon^2)+12\sqrt{6}x\epsilon(y^2\lambda
-\beta\Omega_m))}{2(1-6\epsilon(x^2-1)+108x^2\epsilon^2)}\nonumber\\
&\qquad \qquad \qquad \qquad +\frac{(1+6\epsilon)(3y^2-3-\Omega_r)}{2(1-6\epsilon(x^2-1)+108x^2\epsilon^2)},\\
\frac{\ddot{\pi}}{H\dot{\pi}}&=\frac{36x^3\epsilon+\sqrt{\frac{3}{2}}(y^2\lambda-\beta\Omega_m)
+3\sqrt{6}x^2\epsilon(\beta\Omega_m-y^2\lambda)}{x+6x\epsilon+6x^3\epsilon(18\epsilon-1)}\nonumber\\
&\qquad \qquad \qquad \qquad -\frac{3x(1+6y^2\epsilon-2\epsilon\Omega_r)}{x+6x\epsilon+6x^3\epsilon(18\epsilon-1)},\\
\Omega_m &=1-(x^2(1+18\epsilon)+y^2+\Omega_r).
\end{align}

In our analysis we consider the the exponential form of potential:
\begin{align}
V(\pi)= V_0 e^{\frac{-\lambda \pi}{M_{\rm pl}}}.
\label{pot}
\end{align}

In this case we have, $\Gamma=1$ and therefore $\lambda$ is a constant.
Therefore the system reduces to the set of four autonomous equations.
From the autonomous system, we notice that, when $\epsilon=0$, the model is similar to the standard coupled dark-energy 
model \cite{Amendola:1999er}. The model has the same dynamical phase plane as that of  coupled quintessence except one additional de-Sitter
solution, for $(\epsilon=-1/6,\lambda=0,\Omega_r=0,y^2-2x^2=1)$. 
This solution exists only for the exponential form of potential. Numerical analysis of the autonomous system also depicts that the
cosmology of the model is similar to the case of coupled
quintessence. From Fig. \ref{Fig1} and Fig. \ref{Fig2}, it is evident that, the influence of the
field coupling to $G^{\mu\nu}$  disappears fast with evolution leaving behind, the coupled quintessence scenario and providing  observed cosmic 
acceleration at late times. The plots show tracker behavior of the field. The coupling constant $\beta$ controls the matter phase. 
The parameters of the model can be fixed conveniently to produce a viable late time cosmology. We notice, that the successive sequence 
of radiation, matter and dark energy epochs are obtained in this model.


\section{Cosmological Perturbation}

In this section, we analyse the cosmological perturbation of slotheon scalar field in Newtonian gauge.
The scalar perturbed metric is given by,

\begin{eqnarray}
 ds^2 &=& -(1+2\Psi)dt^2 + a(t)^2 (1-2\Phi)\gamma_{ij}dx^i dx^j\,.
\end{eqnarray}

The equation for linear matter perturbation in the sub-horizon approximation is given by,

\begin{eqnarray}
 \ddot{\delta}_m + (2H+ \frac{\beta}{M_{\rm pl}} \dot\phi)\dot{\delta}_m - 4\pi G_{\rm eff} \rho_m \delta_m &=&0\,\,,
\label{matpert}
 \end{eqnarray}
where,
\begin{widetext}
\begin{align}
G_{\rm eff}=& G\Bigl[1+\frac{2\beta^2(1-6x^2 \epsilon)^2+
2x\epsilon(9x+4\sqrt{6}(\frac{\ddot{\pi}}{H \dot{\pi}})\beta)
-12x^2 \epsilon^2(-9-6\frac{\dot{H}}{H^2}-4(\frac{\ddot{\pi}}{H \dot{\pi}})^2+3x^2+4\sqrt{6}(2+\frac{\ddot{\pi}}{H \dot{\pi}})x\beta)
}
{1+2\epsilon(3+2\frac{\dot{H}}{H^2}-6x^2)+12x^2\epsilon^2(3x^2+8\frac{\ddot{\pi}}{H \dot{\pi}}-4\frac{\dot{H}}{H^2}-2)
+72x^4\epsilon^3(2\frac{\dot{H}}{H^2}-8\frac{\ddot{\pi}}{H \dot{\pi}}-9)}\nonumber\\
&-\frac{72x^4\epsilon^3(-9+2\frac{\dot{H}}{H^2}-8\frac{\ddot{\pi}}{H \dot{\pi}})}
{1+2\epsilon(3+2\frac{\dot{H}}{H^2}-6x^2)+12x^2\epsilon^2(3x^2+8\frac{\ddot{\pi}}{H \dot{\pi}}-4\frac{\dot{H}}{H^2}-2)
+72x^4\epsilon^3(2\frac{\dot{H}}{H^2}-8\frac{\ddot{\pi}}{H \dot{\pi}}-9)}\Bigr]\,, 
\end{align}
\end{widetext}
$\delta_m$ is the gauge invariant density contrast given by $\delta_m=\frac{\delta\rho_m}{\rho_m}+3Hv$,
$v$ being the peculiar velocity of fluid.

\begin{figure}[t!]
\centering
\includegraphics[scale=.6]{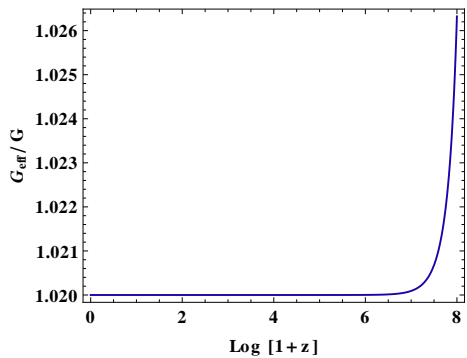}
\caption{Evolution of $\frac{G_{eff}}{G}$ with $\beta=0.1$.}
\label{plotGeff}
\end{figure}

 We show the evolution of $G_{\rm eff}$ as a function of redshift, in Fig. \ref{plotGeff}. It is evident from the 
figure that, because of large gravitational strength in the past, we have
 $G_\text{eff}$ different to General Relativity at large redshifts, whereas $G_\text{eff}=G(1+2\beta^2)>G$ in
 the de Sitter phase. Thus the model reduces to
 coupled quintessence scenario at large scales thereby giving rise to strong cosmological
 constraints on $\beta$ as we shall see in the observational analysis of the model.

\section{Data Analysis}

This is the era of precision cosmology. Astrophysical observations can provide precise data to constrain the
models of dark energy. In this work we use growth, Type Ia Supernova (SNIa), Baryon Acustic Oscillation (BAO) and Cosmic Microwave
Background (CMB) data.

We have used the growth data from the references \cite{growth}. The growth factor $f$ is defined as:

\begin{equation}
 f=\frac{{\rm d} \ln \delta_m}{{\rm d} \ln a}\,.
 \label{grth}
\end{equation}
One can express the Eq.(\ref{matpert}) in terms of the growth factor as,
\begin{align}
\frac{df}{dN}+f^2+f\left(\frac{1}{2}-\frac{3}{2}w_{\rm eff}+ \beta \frac{\dot{\pi}}{H M_{\rm pl}}\right)
- \frac{3}{2} \frac{G_{\rm eff}}{G} \Omega_m =0\,.
\end{align}

\begin{figure}[t!]
\centering
\includegraphics[scale=.48]{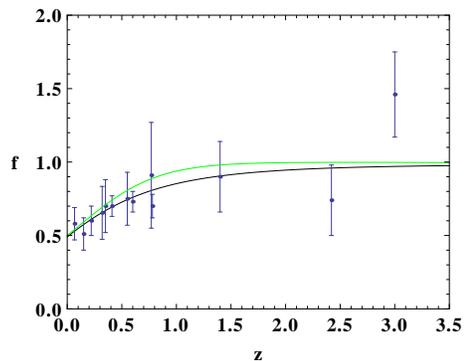}
\caption{Evolution of the growth factor $f$ for the potential studied.
The integration is done from $\Omega_m\approx 0.90$  with $\beta=0.01$. The green line corresponds to $\Lambda$CDM model.}
\label{growth} 
\end{figure}

In Fig. \ref{growth}, we show the evolution of the growth factor $f$ for the exponential
  potential. In this case,
  the deviation from $\Lambda$CDM model is not significant. We also notice that $f< 1$ for all redshift which means
growth is in accordance with the Einstein-de-Sitter model. This goes well with the prediction of Dark Energy scenario.

We define
\begin{align}
\chi^2=\chi_{Growth}^2+\chi_{SN}^2+\chi_{BAO}^2+\chi_{CMB}^2.
\end{align}
 $\chi_{Growth}^2$ is defined as,

\begin{align}
\chi_{Growth}^2(\theta)=\sum_i \frac{f_{obs}(z_i)-f_{th}(zi,\theta)}{\sigma_f(z_i)}\,,
\end{align}
where, $f_{\text{obs}}$ is the observational growth data, $f_{th}$ is the theoretically calculated value of growth factor
of the model from the Eq.$(\ref{grth})$, $\theta$ is the model parameter
 and $\sigma_f$ is the 1$\sigma$ error.
For the Supernova type Ia data, we use the latest UNION2.1 compilation  \cite{Suzuki:Union2.1} which contains 580 data points.
 The data is available in the form of distance modulus $\mu$ with the redshift $z$.
The distance modulus $\mu$ is defined as,
\begin{align}
 \mu=m - M=5 \log D_L+\mu_0\,,
\end{align}
 where $m$ and $M$ are the apparent and absolute magnitudes of the Supernovae respectively, $D_L$  is the luminosity distance defined as
\begin{align}
D_L(z)=(1+z) \int_0^z\frac{H_0dz'}{H(z')}\,,
\end{align}
 and $\mu_0=5 \log\left(\frac{H_0^{-1}}{M_{pc}}\right)+2
5$ is a nuisance parameter which should be marginalized. The corresponding $\chi^2$ is defined as,

\begin{align}
\chi_{SN}^2(\mu_0,\theta)=\sum_{i=1}^{580} \frac{\left(\mu_{th}(z_i,\mu_0,\theta)-\mu_{obs}(z_i)\right)^2}{\sigma_\mu(z_i)^2}\,,
\end{align}
where, $\mu_{obs} $ is the observational distance modulus, $\mu_{th}$ is the theoretical distance modulus of the model 
and $\sigma_{\mu}$ is the error in the distance modulus. Marginalizing the nuisance parameter $\mu_0$ as \cite{Lazkoz:2005sp} we obtain,

\begin{align}
\chi_{SN}^2(\theta)=A-\frac{B^2}{C}\,.
\end{align}

Where,
\begin{align}
&A(\theta) =\sum_{i=1}^{580} \frac{\left(\mu_{th}(z_i,\mu_0,\theta)-\mu_{obs}(z_i)\right)^2}{\sigma_\mu(z_i)^2},\\
&B(\theta) =\sum_{i=1}^{580} \frac{\mu_{th}(z_i,\mu_0,\theta)-\mu_{obs}(z_i)}{\sigma_\mu(z_i)^2},\\
&C(\theta) =\sum_{i=1}^{580} \frac{1}{\sigma_\mu(z_i)^2}.
\end{align}

We have used the BAO data of $\frac{d_A(z_\star)}{D_V(Z_{BAO})}$ \cite{BAO1,BAO2,BAO3,BAO4}, where $z_\star$ is the
 decoupling time $z_\star \approx 1091$, $d_A$ is the comoving angular-diameter distance given by $d_A(z)=\int_0^z \frac{dz'}{H(z')}$ 
and $D_V(z)=\left(d_A(z)^2\frac{z}{H(z)}\right)^{\frac{1}{3}}$. Data required for
this analysis is depicted in table $\ref{baodata}$. We calculate  $\chi_{BAO}^2$ as described in Ref. \cite{BAO5}, where it is defined as,

\begin{equation}
 \chi_{BAO}^2=X_{BAO}^T C_{BAO}^{-1} X_{BAO}.
\end{equation}
Where,
\begin{equation}
X_{BAO}=\left( \begin{array}{c}
        \frac{d_A(z_\star)}{D_V(0.106)} - 30.95 \\
        \frac{d_A(z_\star)}{D_V(0.2)} - 17.55 \\
        \frac{d_A(z_\star)}{D_V(0.35)} - 10.11 \\
        \frac{d_A(z_\star)}{D_V(0.44)} - 8.44 \\
        \frac{d_A(z_\star)}{D_V(0.6)} - 6.69 \\
        \frac{d_A(z_\star)}{D_V(0.73)} - 5.45
        \end{array} \right)
\end{equation}

and the inverse covariance matrix,

\begin{widetext}
\begin{align}
C^{-1}=\left(
\begin{array}{cccccc}
 0.48435 & -0.101383 & -0.164945 & -0.0305703 & -0.097874 & -0.106738 \\
 -0.101383 & 3.2882 & -2.45497 & -0.0787898 & -0.252254 & -0.2751 \\
 -0.164945 & -2.45499 & 9.55916 & -0.128187 & -0.410404 & -0.447574 \\
 -0.0305703 & -0.0787898 & -0.128187 & 2.78728 & -2.75632 & 1.16437 \\
 -0.097874 & -0.252254 & -0.410404 & -2.75632 & 14.9245 & -7.32441 \\
 -0.106738 & -0.2751 & -0.447574 & 1.16437 & -7.32441 & 14.5022
\end{array}
\right).
\end{align}
\end{widetext}

Finally for the constraints from CMB we have used the CMB shift parameter $R=H_0 \sqrt{\Omega_{m0}} \int_0^{1089}\frac{dz'}{H(z')}$, where $\Omega_{m0} $ is the 
present density parameter of matter and $H_0$ is the present Hubble parameter. The $\chi_{CMB}^2$ is defined as,

\begin{align}
 \chi_{CMB}^2(\theta)=\frac{(R(\theta)-R_0)^2}{\sigma^2}
\end{align}

Where, $R_0=1.725 \pm 0.018$ \cite{CMB}.

\begin{figure}[h]
\centering
\includegraphics[scale=.6]{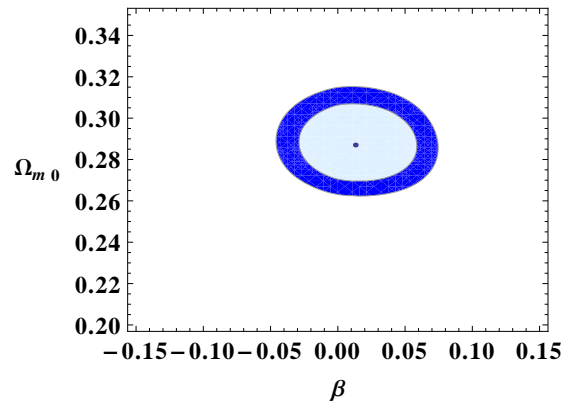}
\caption{The $1\sigma $(light) and $2\sigma$ (dark) likelihood contours
in the ($\Omega_{m0}$ , $\beta$) phase plane for total $\chi^2_{Growth+SN+BAO+CMB }$. The point in the centre
depicts the best fit values of ($\Omega_{m0}$ , $\beta$).
}
\label{chitot} 
\end{figure}

\begin{figure}[h]
\centering
\includegraphics[scale=.6]{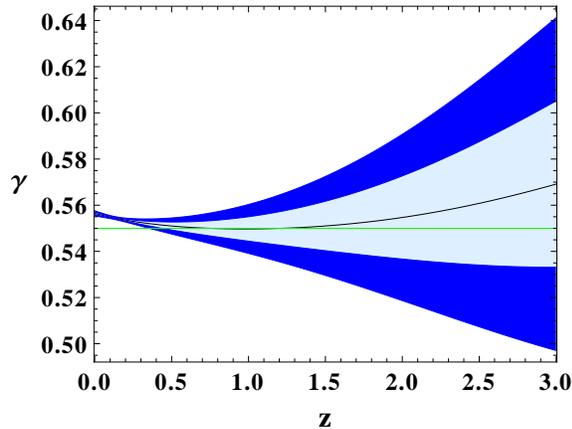}
\caption{The $1\sigma $(light) and $2\sigma$ (dark) of the growth index.
The central line is the growth index for the best fit values
of the model parameters $\beta$ and $\Omega_{m0}$. The green line corresponds to $\Lambda$CDM model.
}
\label{growthindex} 
\end{figure}

We have carried out this analysis on two model parameters $\beta$ and $\Omega_{m0}$.
 To examine the constraints on $\beta$ and $\Omega_{m0}$ we  varied $\beta$ from $-0.3$
 to $1$ and $\Omega_{m0}$ from $0.2$ to $0.35$. 
Fig. \ref{chitot} shows the $1\sigma $ and $2\sigma$ contours in the $(\beta, \Omega_{m0})$ parameter space.
 The total $\chi^2$ minimum is at $\beta \sim 0.014$ and $\Omega_{m0} \sim 0.287$. 
We use these best fit values of the model parameters to plot the growth index $\gamma$ 
in  Fig. \ref{growthindex} with the $1\sigma$ and $2\sigma$ errors.
 We notice that at late time, when  $w>-1$, the evolution is consistent with Dark Energy models  \cite{Polarski:2007rr}. 
But the value of $\gamma$ of the model at  all redshift is large compared to that of $\Lambda$CDM. This is a unique characteristic of this model as it is different in case of
 $f(R)$-gravity models \cite{Gannouji:2008wt} 
or scalar-tensor theories \cite{Gannouji:2008jr} where $\gamma<0.55$.

\begin{center}
\begin{table*}
\begin{tabular}{|c||c|c|c|c|c|c|}
\hline
 $z_{BAO}$  & 0.106  & 0.2 & 0.35 & 0.44 & 0.6 & 0.73\\
\hline \hline
 $\frac{d_A(z_\star)}{D_V(Z_{BAO})}$ &  $30.95 \pm 1.46$ & $17.55 \pm 0.60$ & $10.11 \pm 0.37$ & $8.44 \pm 0.67$ & $6.69 \pm 0.33$ & $5.45 \pm 0.31$  \\
\hline
\end{tabular}
\caption{Values of $\frac{d_A(z_\star)}{D_V(Z_{BAO})}$ for different values of $z_{BAO}$.}
\label{baodata}
\end{table*}
\end{center}
\section{Conclusion}

In this paper we have investigated the slotheon gravity model with a potential.
 The slotheon gravity in general respects the galileon symmetry
in curved space time. Here the field $\pi$ has less kinetic energy than the canonical scalar field.
Though adding a potential breaks the symmetry, but it serves an important role in obtaining a viable cosmology. We have studied the model taking the exponential 
form of the potential. We have demonstrated that the model gives an accelerating universe at late times.
 We found that the cosmology of the model is similar to the case of coupled
quintessence at late times thereby giving current cosmic acceleration.

We have investigated the perturbation of this model. The deviation of the growth factor compared to
 $\Lambda$CDM model is negligible. The growth is in accordance with the Einstein-de-Sitter model. In the case of pure
 coupled quintessence, we have  $G_{\rm eff}=G(1+2\beta^2)$ which requires the coupling
 to be small in order to respect the BBN constraint. On the other
 hand, it is interesting to note that in the model under consideration, $G_{\rm eff}$ is large in the higher redshift than 
 in General Relativity which certainly gives different BBN constraints.

We have also imposed observational constraints on the model parameters using  the data from growth, SNIa,
BAO, and CMB observations. 
Constructing the corresponding contour plots we deduced that  $\beta$ is constrained by the data to small
 range of values $-0.05 < \beta < 0.08$ and the density of matter today is constrained around
the concordance values. The model under consideration shows  larger value of growth index
 for all redshift compared to $\Lambda$CDM.

\section{ACKNOWLEDGEMENTS}
A.A thanks Radouane Gannouji for useful discussions. D.A thanks Atanu Kumar for discussions on the cosmological perturbation theory.

\end{document}